\documentclass[]{interact}
\usepackage{amsmath}
\usepackage{amssymb}
\usepackage{graphicx}
\usepackage{epstopdf}
\usepackage{subfigure}
\usepackage{hyperref}
\usepackage{natbib}
\usepackage{xcolor}
\bibpunct[, ]{(}{)}{;}{a}{}{,}
\renewcommand\bibfont{\fontsize{10}{12}\selectfont}

\theoremstyle{plain}

\theoremstyle{definition}

\theoremstyle{remark}

\begin{document}

\articletype{ARTICLE}

\title{A comparative study of 
non-deep learning, deep learning, and ensemble learning methods for sunspot number prediction}

\author{
\name{Yuchen Dang\textsuperscript{a}, Ziqi Chen\textsuperscript{b}, 
Heng Li\textsuperscript{c}
and Hai Shu\textsuperscript{d,*}\thanks{*CONTACT Hai Shu. Email: hs120@nyu.edu}}
\affil{\textsuperscript{a}Center for Data Science, New York University, New York, NY, USA}
\affil{\textsuperscript{b}Key Laboratory of Advanced Theory and Application in Statistics and Data Science-MOE,
School of Statistics, East China Normal University, Shanghai, China}
\affil{\textsuperscript{c}
Department of Computer Science and Engineering, Southern University of Science and Technology, Shenzhen, Guangdong Province, China
}
\affil{\textsuperscript{d}Department of Biostatistics, School of Global Public Health, New York University, New York, NY, USA}
}

\maketitle

\begin{abstract}
Solar activity has significant impacts on human activities and health.
One most commonly used measure of solar activity is the sunspot number. This paper compares three important non-deep learning models, four popular deep learning models,
and their five ensemble models
in forecasting sunspot numbers. 
{\color{black}In particular, we propose an ensemble model called XGBoost-DL, which uses XGBoost 
as a two-level nonlinear ensemble method
to combine the deep learning models.
Our XGBoost-DL
achieves the best forecasting performance 
(RMSE $=25.70$ and MAE $=19.82$) in the comparison,
outperforming the best non-deep learning model SARIMA (RMSE $=54.11$ and MAE $=45.51$), the best deep learning model Informer (RMSE $=29.90$ and MAE $=22.35$)
and the NASA's forecast (RMSE $=48.38$ and MAE $=38.45$).}
Our XGBoost-DL forecasts a peak sunspot number of 133.47 in May 2025 for Solar Cycle 25 and 164.62 in November 2035 for Solar Cycle 26, similar to but later than the NASA's at 137.7 in October 2024 and 161.2 in December~2034.
{\color{black}
An open-source Python package of our XGBoost-DL
for the sunspot number prediction
is available at \url{https://github.com/yd1008/ts_ensemble_sunspot}.
}

\end{abstract}


\section{Introduction}

Human activities and various events on Earth are strongly intertwined with solar activity \citep{pulkkinen2007space,hathaway2015solar}. 
An increase in solar activity
includes increases in extreme ultraviolet and X-ray emissions
from the Sun
toward Earth, resulting in 
the atmospheric heating 
that can be harmful to spacecrafts, satellites and radars \textcolor{black}{\citep{walterscheid1989solar,ruohoniemi1997rates,lybekk2012solar}.}
Increased solar flares and coronal mass ejections due to
high solar activity 
can damage the communication and power systems on Earth \textcolor{black}{\citep{lewandowski2015massive}}.
The approximately 11-year cyclic pattern of  solar activity seems easily predictable, but the cycle varies in both amplitude and duration \textcolor{black}{\citep{cameron2017understanding}.} Accurate prediction of solar activity is thus of great interest
to estimate the expected impact of space weather on space missions and societal technologies.

Solar cycles are also considered to impact many aspects of human health.
\citet{juckett1993correlation} observed longer human longevities during solar cycle minimums.
\citet{davis2004chaotic} reported higher incidences of mental illness in chaotic solar cycles.
\citet{azcarate2016influence} found larger fluctuations in blood pressure during ascending phases of solar cycles. 
\citet{qu2016sunspot} concluded that most influenza pandemics occurred within one year of solar cycle peaks.
Predictions of solar activity can hence assist people in taking the necessary precautions.

The number of the sunspots, which appear as dark areas on the solar disk, is the most commonly used measure of solar activity \citep{usoskin2017history}. For one thing, the sunspot number is one directly visible characteristic of the Sun. For another, there is a publicly available record of sunspot numbers that can be traced back to as early as year~1749.

Prediction of sunspot numbers
belongs to the scope of time-series forecasting that
may be tackled by either non-deep learning or deep learning methods.
Examples of popular non-deep learning methods in general time series forecasting are
Auto-Regressive Moving Average (ARMA) models \citep{Box1976}, 
 Exponential Smoothing \citep{Holt1957,winters1960forecasting}, and the Prophet model recently released by Facebook \citep{taylor2018forecasting}. Deep-learning time-series forecasting methods, mainly prevailing in natural language processing, 
include Long Short-Term Memory \citep[LSTM;][]{hochreiter1997long}, Gated Recurrent Unit \citep[GRU;][]{cho2014,chung2014empirical}, Transformer \citep{vaswani2017attention}, and the recent Informer \citep{zhou2021informer}. Non-deep learning methods often have restrictive theoretical assumptions that limit their performance on real-world time series data \citep{lara2021experimental}. Deep learning methods are generally superior over non-deep learning methods due to the capability of extracting complex data representations at high levels of abstraction \citep{han2019review}.

Although there have been many studies on predicting the sunspot number
by using non-deep learning \citep{xu2008long, hiremath2008prediction,chattopadhyay2011trend,tabassum2020approach} or deep learning forecasting methods \citep{pala2019forecasting,benson2020forecasting,arfianti2021sunspot,prasad2022prediction},
most are based on ARMA models or deep learning methods
like LSTM or GRU. 
Little work has been done on the 
more recent time-series models Prophet, Transformer, and Informer
for the sunspot number prediction.
Moreover, the ensemble learning methods \citep{kuncheva2014combining} are rarely considered in this forecasting problem.

Ensemble learning is widely used in machine learning to boost the performance by combining results from multiple models. Since  the ensemble result is often better than that of a single model, this technique has been increasingly applied to time series forecasting \citep{wichard2004time,qiu2014ensemble,oliveira2015ensembles,kaushik2020ai}. Basic ensemble methods simply use the average or median of predictions from all base models. More sophisticate methods assign different weights to base models such as 
the error-based method and linear regression \citep{adhikari2014performance}.

Due to the regression nature of ensemble learning, any machine learning algorithm for regression can be used as an ensemble method.
We thus propose to apply one state-of-the-art nonlinear method,
Extreme Gradient Boosting \citep[XGBoost;][]{chen2016xgboost},
to ensemble the predictions from multiple time-series forecasting models.
XGBoost is a fast and accurate implementation of
gradient boosted decision trees, which has recently been dominating applied machine learning for regression and classification tasks \citep{zhong2018xgbfemf,chang2018application,pan2018application,ogunleye2019xgboost}.
The popularity of XGBoost mainly stems from its three modifications to traditional gradient boosted decision trees: the approximate greedy algorithm with weighted quantile sketch to fit large datasets, the sparsity-aware split finding algorithm to deal with missing values, and the cache-aware access technique to effectively utilize hardware resources. 
Besides these advantages, we essentially use 
XGBoost as a two-level ensemble method,
because
XGBoost itself is an ensemble of decision trees and each decision tree therein nonlinearly combines the forecasting models.
The two-level nonlinear ensemble nature of our usage of XGBoost makes it potentially more powerful than single-level ensemble methods such as those aforementioned.

In this paper, we conduct a comparative study of non-deep learning and deep learning models
as well as their ensemble models for the sunspot number prediction.
We compare the three important non-deep learning models, Seasonal Autoregressive Integrated Moving Average \citep[SARIMA;][]{Box1976},
Exponential Smoothing, and Prophet,
and the four popular deep learning models, LSTM, GRU, Transformer, and Informer.
We also consider their ensemble models from
basic ensembles, 
the error-based method,
linear regression,
and XGBoost.

The contributions of the paper are summarized below. 
\begin{itemize}
    \item We compare three important non-deep learning models (SARIMA, Exponential Smoothing, and Prophet), four popular deep learning models (LSTM, GRU, Transformer, and Informer), 
and their five ensemble models
(via mean, median, the error-based method, linear regression, and XGBoost)
in predicting the sunspot number. 

    \item We propose to use XGBoost as a two-level nonlinear ensemble method to combine the results from time-series forecasting models.
Our XGBoost-DL model, 
which uses XGBoost to ensemble
the four deep learning models,
has the best performance in comparison with other considered base and ensemble models as well as 
the prediction from
the National Aeronautics and Space Administration (NASA).

    \item We provide 
an open-source Python package of the XGBoost-DL model for the sunspot number prediction
at \url{https://github.com/yd1008/ts_ensemble_sunspot}.

    \item We 
use the proposed XGBoost-DL model to     
    forecast the Solar Cycles 25 and 26, and compare the result with the NASA's prediction.

\end{itemize}

The rest of this paper is organized as follows.
Section~\ref{sec: methods} introduces the seven aforementioned time-series forecasting methods.
Section~\ref{sec: Ensemble} describes the five ensemble learning methods including the proposed XGBoost-based ensemble.
Section~\ref{sec: Experiments} compares all considered base and ensemble models as well as NASA's report for the sunspot number prediction.
Section~\ref{sec: Conclusion}
makes concluding remarks.

\section{Forecasting Methods}\label{sec: methods}
In this section, we introduce 
seven time-series forecasting methods
{\color{black}that will be compared in our experiments
for predicting sunspot numbers,}
including the three non-deep learning methods
SARIMA, Exponential Smoothing, and Prophet, and the four deep learning
methods LSTM, GRU, Transformer, and Informer.
{\color{black}
SARIMA and Exponential Smoothing
are classical non-deep learning methods
widely used for decades 
in forecasting seasonal time series \citep{Box1976,Holt1957,winters1960forecasting}.
Facebook's Prophet is relatively new compared to the former two but has been used internally in Facebook for years and also allows for seasonality \citep{taylor2018forecasting}.
LSTM and its simpler variant GRU
are two most popular deep-learning models based on gating mechanisms to tackle sequential prediction \citep{hochreiter1997long,cho2014,chung2014empirical}.
Transformer is 
an innovative deep-learning network
instead using a self-attention mechanism \citep{vaswani2017attention},
and its variant Informer, the winner of AAAI-21 Outstanding Paper Award,
improves Transformer on long-sequence time series forecasting \citep{zhou2021informer}.
We are aware of many other models \citep[e.g.,][]{tabassum2020approach,benson2020forecasting,beltagy2020longformer,child2019generating}, most of which are similar to 
or variants of the above seven methods, 
but we only consider
these seven methods that are
most widely recognized in time series forecasting.
}

We denote a univariate time series by 
$y = (y_1, y_2, \dots, y_t, \dots)$, where $y_t$ is the observation at time $t$. 

\subsection{Non-deep learning methods}
\subsubsection{SARIMA}
SARIMA \citep{Box1976}, as an ARMA variant, is one most commonly used model in the past decades to forecast
trend and seasonal time series.
It uses a mix of autoregressive terms, moving average terms, and differencing procedures for both non-seasonal and seasonal components to represent the current value in a time series based on prior observations.
Specifically, an SARIMA$(p,d,q)(P,D,Q)_s$ model is defined by
\begin{align*}
    \phi(L)\Phi(L^s)(1-L)^d(1-L^s)^D y_t &= c+\theta(L)\Theta(L^s)\epsilon_t  \\
    \phi(L) &= 1-\phi_1 L^1-\phi_2 L^2 - \dots - \phi_p L^p\nonumber\\
    \Phi(L^s) &= 1-\Phi_1 L^s - \Phi_2 L^{2s} - \dots - \Phi_P L^{Ps}\nonumber\\
    \theta(L) &= 1+\theta_1 L^1 +\theta_2 L^2 + \dots + \theta_q L^q\nonumber\\
    \Theta(L^s) &= 1+\Theta_1 L^s +\Theta_2 L^{2s} + \dots + \Theta_Q L^{Qs}, \nonumber
\end{align*}
where $(\phi_i,\theta_i,\Phi_i,\Theta_i)$ and $(p,q,P,Q)$ are the parameters and the orders of the non-seasonal autoregressive, the non-seasonal moving average, the seasonal autoregressive,  and the seasonal moving average terms, respectively, $\epsilon_t$ is white noise, $L$ is the lag operator,  $d$ is the order of non-seasonal differencing,  $D$ is the order of seasonal differencing, $s$ is the span of the seasonality, and $c$ is a constant. 

\subsubsection{Exponential Smoothing}
Proposed in 1950s, Exponential Smoothing \citep{brown1956exponential, Holt1957,winters1960forecasting} remains to be one of the widely used time series forecasting methods. 
Although there are several types of Exponential Smoothing, we focus on the Holt-Winters Exponential Smoothing method \citep{Holt1957,winters1960forecasting}, which can model the trend and seasonality of a time series.
The Holt-Winters method and SARIMA
have shown comparable performances in a number of previous studies \citep{lidiema2017modelling, liu2020forecast, rabbani2021comparison}. 

Also known as Triple Exponential Smoothing, the Holt-Winters method comprises three smoothing equations for the level $l_t$, the trend $b_t$, and the seasonality $s_t$, respectively. 
There are two main models of this method, the additive seasonal model 
\begin{align*} 
    y_{t+h} &= l_t + hb_t+s_{t-m+h_m^+} \\ 
    l_t &= \alpha (y_t - s_{t-m}) + (1-\alpha)(l_{t-1}+b_{t-1}) \nonumber \\
    b_t &= \beta(l_t - l_{t-1}) + (1-\beta)b_{t-1} \nonumber\\
    s_t &= \gamma(y_t-l_{t-1}-b_{t-1})+(1-\gamma)s_{t-m}, \nonumber
\end{align*}  
and the
multiplicative seasonal model
\begin{align*} 
    y_{t+h} &= (l_t + hb_t)s_{t-m+h_m^+}\\ 
    l_t &= \alpha \frac{y_t}{s_{t-m}} + (1-\alpha)(l_{t-1}+b_{t-1}) \nonumber \\
    b_t &= \beta(l_t - l_{t-1}) + (1-\beta)b_{t-1} \nonumber\\
    s_t &= \gamma \frac{y_t}{l_{t-1}+b_{t-1}}+(1-\gamma)s_{t-m} ,
    \nonumber
\end{align*}  
where $\alpha,\beta,\gamma \in[0,1]$ are smoothing parameters, 
$m$ is the frequency of the seasonality, and $h_m^+=[(h-1)~\text{mod}~m]+1$ for $h\ge 1$.

The two models differ in the nature of the seasonality.
The additive model ought to be considered when the seasonal variations are stable over time, while the multiplicative model is used when the seasonal variations are changing proportional to the level of the time series.  
Due to the variability of the amplitude of sunspot cycles, following \citet{tabassum2020approach}, we use the multiplicative model for the sunspot number prediction.

\subsubsection{Prophet} 
Facebook's Prophet \citep{taylor2018forecasting} is a more recent time series forecasting algorithm compared to the previous two. Despite some commonalities with SARIMA and Exponential Smoothing, it provides a more intuitive approach to model the trend and seasonality of time series by incorporating more flexibilities in its configuration.

Prophet has three essential components: trend $b_t$, seasonality $s_t$, and holiday $h_t$. 
The holiday option allows Prophet to adjust the forecast that may be affected by holidays or major events.
The full model with a logistic trend term is 
\begin{align*}
    y_t &= b_t+s_t+h_t+\epsilon_t \\
    b_t &= \frac{C}{1+\exp(-k(t-m))} \nonumber \\
    s_t &= \sum_{n=1}^{N}\left[\alpha_n \cos\left(\frac{2\pi nt}{P}\right)+\beta_n \sin\left(\frac{2\pi nt}{P}\right)\right]\nonumber\\
    h_t &= [\mathbf{1}(t \in D_1),\dots,\mathbf{1}(t \in D_L)]\boldsymbol{\kappa},\nonumber
\end{align*}
where $\epsilon_t$ is the error term, 
$C$ is the carrying capacity that is the maximum value of the logistic curve, $k$ is the growth rate that controls the steepness of the curve, $m$ is an offset parameter corresponding to the curve's midpoint, the seasonality $s_t$ is expressed by a standard Fourier series with parameters $\{\alpha_n,\beta_n\}_{n=1}^N$ and a regular period $P$ that the time series has,
$D_i$ is the set of dates for holidays, and $\boldsymbol{\kappa}\in \mathbb{R}^L$ is the change in the forecast caused by holidays.

\subsection{Deep learning methods}
Although non-deep learning methods can handle a wide range of time series forecasting tasks, their theoretical limitations often prevent them from being directly applicable to data or modeling complex non-linearity in data, as well as computational complexity makes them impractical to large datasets. Therefore, deep learning techniques such as Rucurrent Neural Networks (RNNs) were introduced  \citep{lara2021experimental}. 
RNNs have shown better performance than non-deep learning methods
in time series forecasting due to 
the ability to deal with longer sequences and better capture complex temporal dependencies \citep{tsui1995recurrent,zhang1998time}. 
\subsubsection{LSTM}
The LSTM network \citep{hochreiter1997long} is one most popular model in the RNN family. The vanilla RNN suffers from vanishing gradients and is not capable to achieve desirable results for long sequence data \citep{hochreiter1998vanishing}. The LSTM network mitigates this issue rising from long-term dependencies by introducing a gated memory cell architecture,
which controlls the information flow with three gates: an input gate $i_t$, an output gate $o_t$, and a forget gate $f_t$.  
The LSTM cell is formulated as follows:
\begin{align}
    i_t &= \sigma(W_i x_t+U_i h_{t-1}+b_i) \label{eq:lstm} \\
    f_t &= \sigma(W_f x_t+U_f h_{t-1}+b_f) \nonumber \\
    o_t &= \sigma(W_o x_t+U_o h_{t-1}+b_o) \nonumber \\
    \widetilde{c_t} &= \tanh(W_c x_t+ U_c h_{t-1}+b_c) \nonumber \\
    c_t &= f_t \cdot c_{t-1} + i_t \cdot \widetilde{c_t} \nonumber \\
    h_t &= o_t \cdot \tanh(c_t), \nonumber 
\end{align}
where $x_t$ is the input which is the hidden state $h_t$ from the previous layer and is $y_t$ for the first hidden layer,
$\sigma(\cdot)$ is the sigmoid function, $\tanh(\cdot)$ is the hyperbolic tangent function, $h_t$ is the hidden state, $c_t$ is the state of the memory cell, $\widetilde{c_t}$ is the candidate state of the memory cell, and $W$ and $U$ are the weights of the input and recurrent connections as well as $b$ is bias with subscripts $i$, $f$, $o$ and $c$ for the input gate, forget gate, output gate, and memory cell, respectively. 

\begin{figure}[t!]
\centering
\includegraphics[width=10cm]{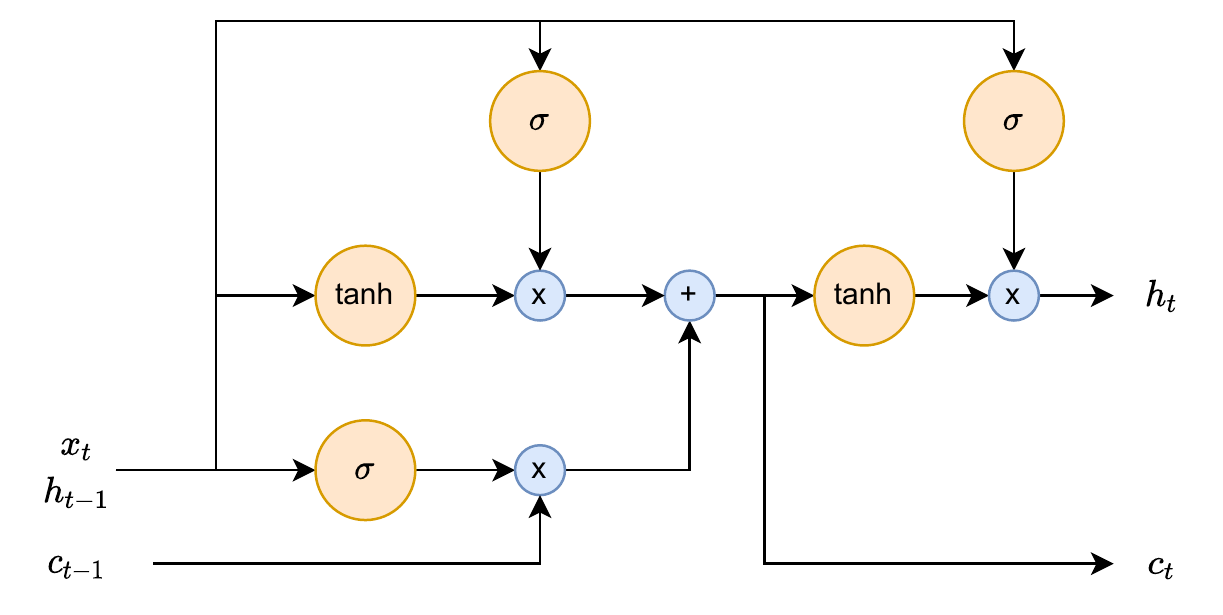}
\caption{The structure of the LSTM cell.} \label{fig:lstm}
\end{figure}

In each hidden layer of the LSTM network, a sequence of LSTM cells are aligned side-by-side and input data are sequentially fed into each cell. LSTM's capability of communicating across multiple cells comes from the hidden states and the cell states. The hidden state carries over information from the previous cell to the next and a new hidden state is generated from each LSTM cell, while the cell state selectively stores the past information. The input, output, and forget gates generate a new hidden state and update the the cell state in the following procedure.
The input gate determines whether new information will be added to the memory in two steps: $i_t$ uses a sigmoid function to decide which information needs to be updated, and $\widetilde{c_t}$ utilizes a hyperbolic tangent function to select new candidate information to be added to the memory.
The forget gate decides what information should be discarded from the memory by applying a sigmoid function to the previous hidden state $h_{t-1}$ and current input value $x_t$. The output of the forget gate ranges between 0 and 1, with 0 indicating complete removal of the previously learnt value and 1 indicating retention of all information.
The output gate determines what will be generated as the output. $o_t$ acts like a filter by selecting the relevant information from the memory with a sigmoid function to generate an output value, which is then multiplied with the cell state passing through a hyperbolic tangent function to form the representations in the hidden state.
The structure of the LSTM cell is illustrated in Figure~\ref{fig:lstm}.

\begin{figure}[b!]
\centering
\includegraphics[width=10cm]{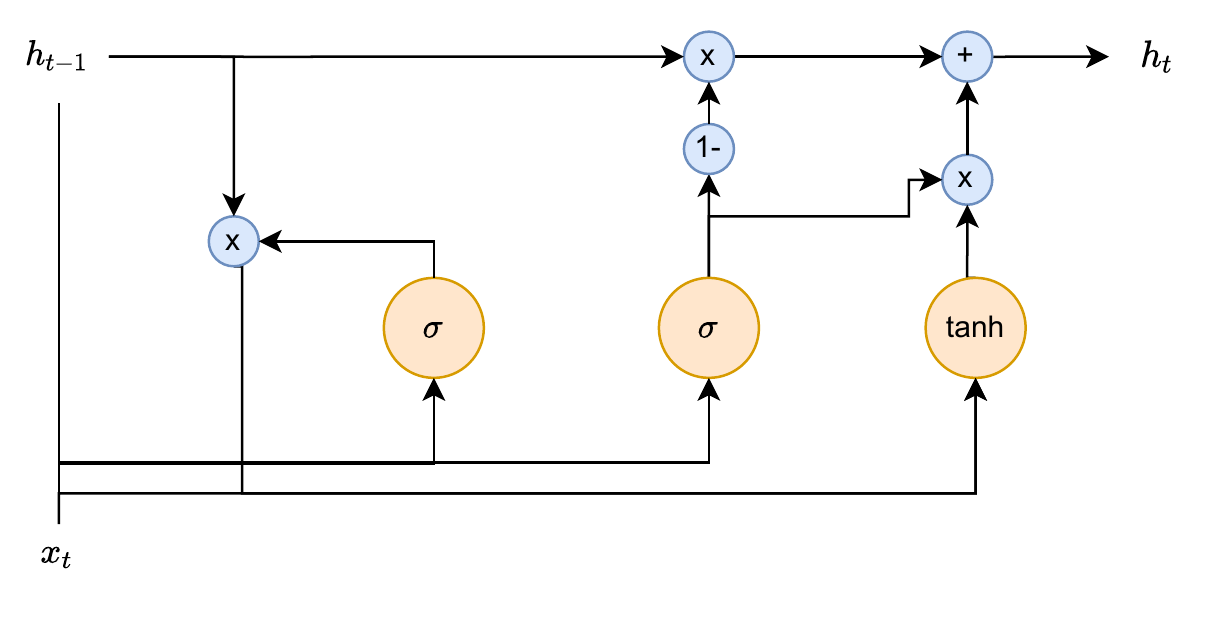}
\caption{The structure of the GRU cell.} \label{fig:gru}
\end{figure}

\subsubsection{GRU}
The GRU \citep{cho2014,chung2014empirical} network is another well-known RNN model using a gating mechanism similar to that in LSTM, but it has a simpler cell architecture and is computationally more efficient \citep{torres2021deep, lara2021experimental}. The GRU cell only has two gates, an update gate $z_t$ and a reset gate $r_t$. The update gate  decides the amount of previous information to be passed to the next state, helping capture long-term dependencies in the sequence. 
The reset gate determines how much of the past information to neglect and is responsible to learn short-term dependencies. 
The GRU cell is formulated by
\begin{align}
    z_t &= \sigma(W_z x_t + U_z h_{t-1} + b_z) \nonumber\\
    r_t &= \sigma(W_r x_t + U_r h_{t-1} + b_r) \nonumber \\
    \widetilde{h_t} &= \tanh(W_h x_t + U_h (r_t \cdot h_{t-1}) + b_h) \nonumber\\
    h_t &= (1-z_t) \cdot h_{t-1} + z_t\cdot \widetilde{h_t}, \nonumber
\end{align}
where the notation is similar to that in \eqref{eq:lstm}. The functionalities of the gates used in GRU are also similar to those in LSTM. Figure \ref{fig:gru} shows the structure of the GRU cell.

\subsubsection{Transformer}
Since introduced in 2017, Transformer \citep{vaswani2017attention},
a solely self-attention based encoder-decoder network, 
has become the state of the art model in natural language processing, along with a number of its variants  \citep{devlin2018bert,yang2019xlnet,liu2019roberta}. 
Recently, Transformer and self-attention based models have gained increasing popularity in time series tasks \citep{wu2020deep,zhou2021informer,wen2020time, li2019enhancing}. 
Transformer abandons the recurrent layers of RNNs that process data of the input sequence one after another.
It instead uses a self-attention mechanism,  
which ditches the sequential operations and can access any part of the sequence, 
to capture global dependencies and enable parallel computation.

\begin{figure}[b!]
\centering
\includegraphics[width=6cm]{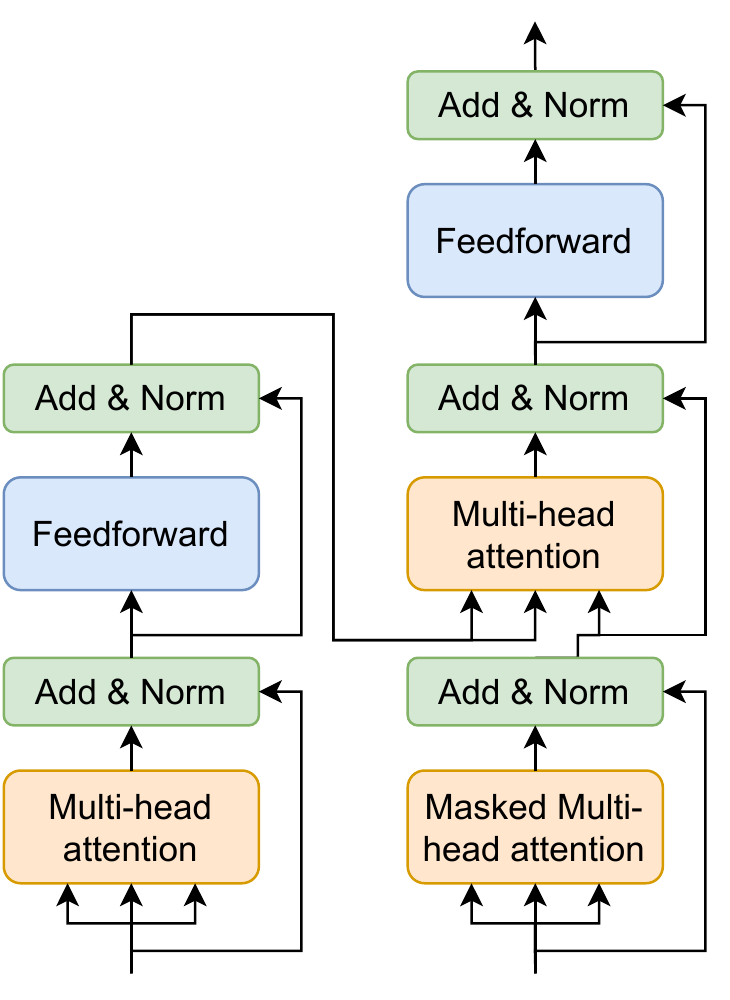}
\caption{Transformer's encoder-decoder architecture. \textcolor{black}{The left four blocks form an encoder layer and the right six blocks form a decoder layer. The encoder's input is first passed through a multi-head attention block, followed by residual connection and layer normalization. The output then goes through a feedforward layer, followed by another residual connection and layer normalization. The decoder's input is fed first into a masked multi-head attention block, then residual connection and layer normalization. The procedure for the next four blocks is the same as for the encoder, except that the encoder output is fed into the multi-head attention block with the decoder output from the second block.}} \label{fig:transformer}
\end{figure}

The core of Transformer is the scaled dot-product self-attention written as
\begin{equation}
    \text{Attention}(Q,K,V) = \text{softmax}(\frac{QK^\top}{\sqrt{d_k}})V,
    \label{eq:selfatt}
\end{equation}
where $Q, K$ and $V$ are matrices with the $i$-th rows being 
the $i$-th query, key and value vectors, and $d_k$ is the dimension of the key vectors.
In a self-attention layer, input data pass through three separate linear layers to form the query, key, and value matrices. Dot products of queries and keys are then calculated. In masked attention layers, a mask of the same size as the dot product matrix, with the upper triangle of the mask having values of $-\infty$ and $0$'s elsewhere, is added to the dot product to prevent values in a sequence from attending to succeeding ones. The dot product matrix scaled by $\sqrt{d_k}$ is then fed into the softmax function. The attention scores are calculated by the dot product between the output of the softmax function and the value matrix. In multi-head attention layers, multiple self-attention layers are stacked with each layer consisting of different sets of weights. The final attention scores are generated by combining attention scores calculated in parallel from each self-attention layer. To use the sequence-order information, Transformer adopts a positional encoding mechanism based on sine and cosine functions. The encoded values are added to the input data, indicating positional information of each value in the sequence, such that Transformer can distinguish values in one position from another without requiring specific order of the input data.
Figure \ref{fig:transformer} illustrates Transformer's encoder-decoder architecture.

\subsubsection{Informer}
The quadratic computational complexity of dot-product self attention and the heavy memory usage in stacking layers are major concerns in dealing with long input sequences. There have been some attempts to address these issues \citep{child2019generating,li2019enhancing, beltagy2020longformer, zhou2021informer}. We focus on the award-winning method Informer \citep{zhou2021informer}. 

Informer is proposed as an enhancement of Transformer on long-sequence time series forecasting. Informer adopts a ProbSparse self-attention mechanism which selects only the most dominant $u$ queries for each key to attend to: 
\begin{equation*}
    \text{Attention}(Q,K,V) = \text{softmax}(\frac{\overline{Q}K^\top}{\sqrt{d_k}})V, 
\end{equation*}
where $\overline{Q}$ only contains the top-$u$ queries chosen under the query sparsity measurement with Kullback-Leibler divergence which is further approximated efficiently with random sampling. The ProbSParse self-attention allows Informer to reduce both time complexity and memory usage from $O(L^2)$ down to $O(L\log L)$ for input length $L$.
In addition, Informer allows longer input sequences by using the self-attention distilling:
\begin{equation*}
    X^{t}_{j+1} = \text{MaxPool}(\text{ELU}(\text{Conv1d}([X_j^t]_{AB}))),
\end{equation*}
where $[\cdot]_{AB}$ is the attention block, $X^{t}_{j+1}$ is the $t$-th sequence in the $j$-th attention layer, MaxPool$(\cdot)$ is a max-pooling layer with stride 2, ELU$(\cdot)$ is the ELU activation function, and Conv1d$(\cdot)$ is a 1-D convolutional filter with kernel size of 3. By distilling, a more condensed feature map is passed from one attention layer to the next, while information is largely preserved. The corresponding memory usage is $O((2-\epsilon)L\log L)$ instead of $O(JL\log L)$ for $J$-stacking layers, where $\epsilon$ is a small number.

\section{Ensemble Learning Methods}\label{sec: Ensemble}
Ensemble learning combines the results from multiple models to produce predictions \citep{dong2020survey}. The combined results usually exhibit better performance because ensemble methods can often decrease the likelihood of overfitting, reduce the chance of being trapped in a local minimum, and extend the size of the search space \citep{sagi2018ensemble}. In time series forecasting tasks, commonly used ensemble methods include basic ensemble methods, error-based method, and machine learning methods. In the following sections, $\widehat{Y}=(\widehat{y}_{ij}) \in \mathbb{R}^{n\times m}$ denotes the matrix of predictions from $m$ forecasting models at $n$ time points, $\widehat{y}_{ij}$ is the predicted value from the $j$-th model at the $i$-th time point, and $\widehat{y}_i$ is the final prediction at the $i$-th time point.

\subsection{Basic ensemble methods}
Basic ensemble methods combine the predictions from all base models by some simple functions such as mean 
and median:
\begin{align*}
\widehat{y}_i&=\text{mean}(\widehat{y}_{i1},\dots, \widehat{y}_{im}),\\
\widehat{y}_i&=\text{median}(\widehat{y}_{i1},\dots, \widehat{y}_{im}).
\end{align*}
 Although these methods are straightforward and simple to comprehend, they ignore the possible relationship among base models and thus are not adequate for 
 combining nonstationary models \citep{allende2017ensemble}. 

\subsection{Error-based method}
Unlike the mean ensemble method that weighs all base models equally, 
error-based method assigns weights inversely proportional to forecast errors \citep{adhikari2014performance}. 
Specifically, the data is divided into training and validation sets. A specific evaluation metric is selected to compute the forecast error $e_j$ on the validation set for the $j$-th base model fitted on the training set, and then the model's weight is 
\begin{equation*}
    w_j = \frac{e_{j}^{-1}}{\sum_{j=1}^{m}e_{j}^{-1}}.
\end{equation*}
The final prediction of this method is
\begin{equation}\label{eq: linear ensemble}
\widehat{y}_{i}=\sum_{j=1}^{m} w_j\widehat{y}_{ij}.
\end{equation}
The error-based method improves the mean ensemble method by allowing different weights for base models, but its performance is still highly dependent on the forecasting performance of each base model. 

\subsection{Linear regression}
The mean, median, and error-based ensemble methods are all linear ensemble methods in the form of \eqref{eq: linear ensemble}.
The linear ensemble problem can be written as the linear model:
\begin{equation}\label{eq: linear model}
y_i=\widehat{y}_i+\varepsilon_i = \sum_{j=1}^{m} w_j\widehat{y}_{ij}
+\varepsilon_i,
\end{equation}
where the forecasts of base models $\{\widehat{y}_{ij}\}_{j=1}^m$ are the features, and the true observation $y_i$ is the target.
Linear regression, or the least squares method, provides the optimal weights that minimize the sum of squared errors $\sum_{i=1}^n\varepsilon_i^2$. 
Formulate \eqref{eq: linear model} in the matrix form
\[
y=\widehat{Y}w+\varepsilon.
\]
The optimal weight vector given by linear regression is $w=\widehat{Y}^\dag y$
with $\widehat{Y}^\dag$ the Moore-Penrose pseudo-inverse of $\widehat{Y}$ \citep{penrose1956best}.

\subsection{XGBoost}
As ensemble learning is essentially a regression problem with features $\{\widehat{y}_{ij}\}_{j=1}^m$ and target $y_i$, any machine learning algorithm for regression is applicable as an ensemble method. In addition to linear regression, there are many nonlinear regression methods such as kernel regression \citep{li2007nonparametric}, support vector regression \citep{cherkassky2004practical}, and tree-based regression algorithms \citep{MR726392, breiman2001random, friedman2001greedy}.

We consider the state-of-the-art nonlinear method, XGBoost \citep{chen2016xgboost}, a highly efficient and effective implementation of gradient boosted decision trees.
XGBoost aims to solve the objective function
\begin{align}
&\mathcal{L}=\sum_{i=1}^n l(\widehat{y}_i, y_i)+\sum_{k=1}^K \Omega (f_k)\label{eq: xgboost}\\
&\text{with} \quad 
\widehat{y}_i=\sum_{k=1}^K f_k({\color{black}x_i})\nonumber\\
&\text{and}\quad 
\Omega(f) = \gamma T+\frac{1}{2}\lambda \|\omega\|_2^2,\nonumber
\end{align}
where $l$ is a differentiable convex loss function
that measures the difference between the prediction $\widehat{y}_i$ and the target $y_i$, 
{\color{black}$x_i$ is the vector of input features,}
$\{f_k\}_{k=1}^K$ are decision trees, and $\Omega(f)$ is the penalty term with tuning parameters $\gamma$ and $\lambda$ to control the model complexity in which $T$ and $\omega$ are the number of leaves and the leaf weights in the tree.
The first term is the objective of the traditional gradient tree boosting, while the second term added by XGBoost is to prevent overfitting. 

{\color{black}To apply XGBoost 
to ensemble 
multiple forecasting models, we simply let the input vector be the predictions of the base forecasting models, that is, $$x_i=(\widehat{y}_{i1},\dots,\widehat{y}_{im}).$$
This is an innovative use of XGBoost,
in contrast to its ordinary usage in
supervised learning where explicit features are ready to be the input. In other words,
XGBoost can not be directly applied to predict the time series  $\{y_i\}_{i\ge 0}$, since we only have
the time series itself and the associated time.
If time is used as the input, then the XGBoost model is approximately equivalent to a high-order polynomial
of time, which is not recommended in time-series literature
because the resulting forecast 
is often unrealistic when the model is extrapolated \citep{hyndman2018forecasting}.
Alternatively, if observations at previous time steps 
are the input, the XGBoost model
resembles the autoregressive model that
only captures short-range temporal dependence \citep{MR1093459}.
Instead of building a XGBoost-based time series model 
from scratch, we
use XGBoost to ensemble 
the predictions from 
existing forecasting methods.


}

We intrinsically use XGBoost as a two-level ensemble method here
to combine the predictions of multiple forecasting methods. 
As shown in \eqref{eq: xgboost}, 
XGBoost itself is an ensemble method that sums the results of $K$ decision trees, each of which is a base model.
In addition to the ensemble nature of XGBoost,
each decision tree therein nonlinearly ensembles the forecasts of base time-series models.
The entire procedure turns out to be an ensemble model that has two levels of ensembles, which is expected to perform better than single-level ensemble methods such as those mentioned above.

{\color{black}
We thus propose a model called XGBoost-DL,
which uses XGBoost to ensemble the
predictions from the four deep learning models
LSTM, GRU, Transformer and Informer. 
Our XGBoost-DL is simple yet novel and effective,
because it takes
advantage of
our innovative use of XGBoost as
a two-level nonlinear ensemble method
and the superior performance
of these deep learning models
in time-series forecasting,
and indeed achieves the best result
in our experiment.
}

\section{Experiments}\label{sec: Experiments}

\subsection{Data description}
The sunspot number dataset is obtained from the website of World Data Center SILSO, Royal Observatory of Belgium, Brussels\footnote{\url{https://www.sidc.be/silso/datafiles}}. The dataset contains 3277 records of monthly averaged total sunspot number from January 1749 to January 2022. 
We also consider NASA's past forecast (from April 1999 to January 2022) and 
most recent forecast (from February 2022 to October 2041) obtained from its website\footnote{\url{https://www.nasa.gov/msfcsolar/archivedforecast}} for comparison with considered methods. \textcolor{black}{
NASA uses a linear regression model based on the 13-month smoothed and
Lagrangian interpolated data to predict sunspot numbers, 
for which a brief description is given in
\citet{suggs2017msfc}
but detailed implementation and computer code are not disclosed. 
}

The observed sunspot number data is split in chronological order into training (2160 records from January 1749 to December 1928), validation (843 records from January 1929 to March 1999), and testing (274 records from April 1999 to January 2022) sets. 
The validation set monitors the training for hyperparameter tuning.
The testing set evaluates the performance of each trained model on unseen data.
{\color{black}The ratio of training and validation sets is 7/3,
and the testing set is concurrent with NASA's past forecast.}
The best model will be used to predict
sunspot numbers from February 2022 to October 2041, covering the remaining portion of
current Solar Cycle~25 and the coming Solar Cycle~26,
in alignment with NASA's current forecasting time range.

 \subsection{Implementation details}
 
 We compare the three non-deep learning models, SARIMA,
Holt-Winters multiplicative Exponential Smoothing, and Prophet,
and the four deep learning models, LSTM, GRU, Transformer, and Informer.
We also consider their ensemble models from
the mean and median ensemble methods,
the error-based method, linear regression,
and our proposed ensemble method via XGBoost.

We adopt the two evaluation metrics,
the root mean squared error (RMSE) and the mean absolute error (MAE), which are widely-used in time series forecasting problems \citep{hyndman2018forecasting}, to assess the prediction performance of considered methods:
\begin{align*}
    \text{RMSE} &= \sqrt{\frac{1}{n}\sum_{i=1}^{n}(\widehat{y}_i-y_i)^{2}}, \\
    \text{MAE} &= \frac{1}{n}\sum_{i=1}^{n}|\widehat{y}_i-y_i|,
\end{align*}
where $\widehat{y}_{i}$ and $y_i$ are the predicted value and the true value, respectively, and $n$ is the number of predictions.

All the experiments are performed in Python 3.8 \citep{10.5555/1593511} environment. All deep learning methods are implemented with Pytorch 1.9.0 \citep{paszke2019pytorch}. SARIMA, basic ensemble methods, and the error-based ensemble method are implemented with Merlion 1.0.0 \citep{bhatnagar2021merlion}. Exponential Smoothing and the linear regression ensemble method are performed with Darts 0.15.0 \citep{herzen2021darts}. The XGBoost ensemble method is carried out using xgboost 1.5.1 \citep{chen2016xgboost}. Models are trained on two NVIDIA RTX8000 GPUs. We use Tune \citep{liaw2018tune} in Ray API 1.9.0 \citep{moritz2018ray} for tuning hyperparameters
{\color{black}with best values chosen as the minimizers of the RMSE on the validation set}.
All deep learning models are trained up to 200 epochs by the AdamW optimizer, and early-stopping is used to prevent overfitting. 
Source codes with pre-trained models, tuning ranges of model-specific hyperparameters, and best configurations for all models used in this study can be found at our GitHub repository\footnote{\url{https://github.com/yd1008/ts\_ensemble\_sunspot}}.

All ensemble methods are applied independently to ensemble the forecast outputs of the three non-deep learning  models, the four deep learning models, and all the seven models, respectively. 
For basic ensemble methods, training is not required and performance is evaluated directly on the testing set. The ensemble weights from the error-based method are calculated by the inverse of the mean squared error (MSE) of each base model on the validation set.
The linear-regression ensemble method
computes the weights from the training and validation sets. 
MSE is chosen as the loss function for training XGBoost-based ensemble models.

\subsection{Results}\label{sec: results}

\begin{figure}
\centering
\subfigure[Forecasts from deep learning models and NASA.]{
\includegraphics[width=\textwidth]{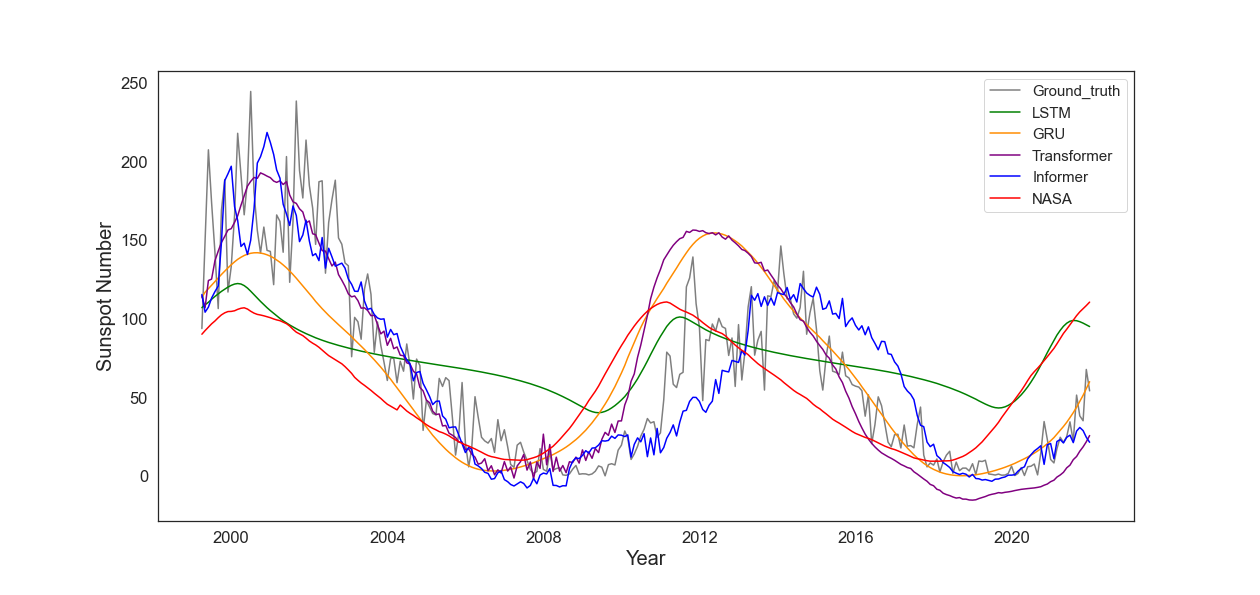}}\\
\subfigure[Forecasts from non-deep learning models.]{
\includegraphics[width=\textwidth]{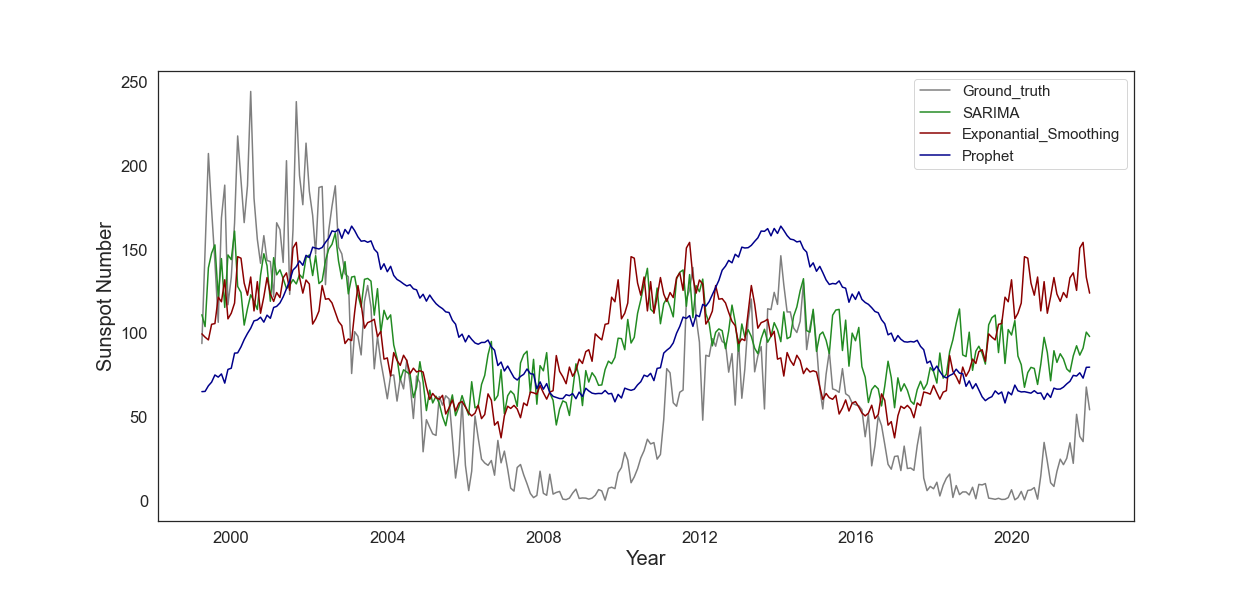}}
\caption{Forecasts of sunspot numbers from base models for the testing set.}
\label{fig:dl_model_pred}
\end{figure}

{\color{black}\subsubsection{In-sample forecast}\label{sec: in-sample forecast}}
Table~\ref{table:performance models} presents the performance results of the seven base models on the testing set. All the four deep learning models outperform the three non-deep learning models. The two attention-based deep learning models (Transformer and Informer) exhibit better performance than the two RNN models (LSTM and GRU). In particular, Informer  achieves the lowest RMSE and MAE with values 29.90 and 22.35, respectively. All the deep learning models except LSTM show more accurate results than NASA with Informer having $38.20\%$ lower RMSE and $41.87\%$ lower MAE. Figure~\ref{fig:dl_model_pred}~(a) displays the true sunspot numbers and their
estimates from deep learning models and NASA for the testing set. It is observed that predictions of deep learning models generally follow along patterns in the ground truth data, while the LSTM and NASA
have large deviations 
around years 2003 and 2020. 
Figure~\ref{fig:dl_model_pred}~(b) shows the forecasts from non-deep learning models.
The three models fail to predict the true trend of the series
{\color{black} due to highly inaccurate forecasts of the peaks and troughs}, albeit have an approximately 11-year cycle. With the lowest RMSE and MAE among the three models, the estimates from SARIMA are rather accurate at the beginning of the testing time horizon but notably deviate from the actual values thereafter. 

\begin{table}[h!]
\tbl{Results of base models on the testing set of sunspot numbers.}
{\begin{tabular}{lcc} \toprule
  & RMSE & MAE  \\ \midrule
 SARIMA & 54.11 & 45.51  \\
 Exponential Smoothing & 61.41 & 49.76 \\
 Prophet & 60.15 & 56.09 \\
 LSTM & 46.14 & 39.44 \\
 GRU & 37.14 & 26.77 \\
 Transformer & 33.99 & 25.56 \\
 Informer & \textbf{29.90} & \textbf{22.35} \\
 NASA & 48.38 & 38.45\\ \bottomrule
\end{tabular}}
\label{table:performance models}
\end{table}

\begin{table}[h!]
\tbl{Results of ensemble models on the testing set of sunspot numbers.}
{\begin{tabular}{lcc|cc|cc} \toprule
 & \multicolumn{2}{c}{Non-deep learning} & \multicolumn{2}{c}{Deep Learning} &\multicolumn{2}{c}{All Models}  \\ \cmidrule{2-7}
  & RMSE & MAE & RMSE & MAE & RMSE & MAE\\ \midrule
Mean & 54.80 & 48.15 & 29.01 & 22.57 & 35.79 & 30.51 \\
  Median & 53.45 & 46.02 & 29.96 & 21.75 & 39.52 & 33.88\\
 Error-based & 54.70 & 48.00 & 27.59 & 21.12 & 31.67 & 26.53 \\
 Regression & 59.98 & 52.21 & 36.26 & 25.72 & 38.54 & 31.04 \\
 XGBoost & 55.45 & 49.58 & \textbf{25.70} & \textbf{19.82} & 30.37 & 22.30\\ \bottomrule
\end{tabular}}
\label{table:performance ensembles}
\end{table}

The improved performance of deep learning methods over non-deep learning ones can be attributed in part to the formers' ability to effectively capture non-linearities, as observed in the complex variations in solar cycles, whereas the latters primarily use linear combinations. Furthermore, the two attention-based methods {\color{black}(Transformer and Informer)} appear to better capture the trend {\color{black}by} using self-attention mechanisms that allow dependencies over a longer period of time to be included into forecasts.

Table \ref{table:performance ensembles} summarizes the testing results from  the ensemble models. 
{\color{black}With larger MAE or RMSE compared to those in Table~\ref{table:performance models},}
all ensembles of non-deep learning models and those of all base models fail to improve the performance 
{\color{black} of SARIMA and Informer, respectively,} since \textcolor{black}{ non-deep learning models are highly inaccurate in predicting the peaks and troughs of the time series as shown in Figure~\ref{fig:dl_model_pred}~(b).}
All ensembles of deep learning models perform better than those of non-deep learning models due to the solid base on the well-performing deep learning models, and ensembles of all base models have performances between those of the former two.

\begin{figure}
\centering
\subfigure[Forecasts from ensembles of deep learning models and from NASA.]{
\includegraphics[width=\textwidth]{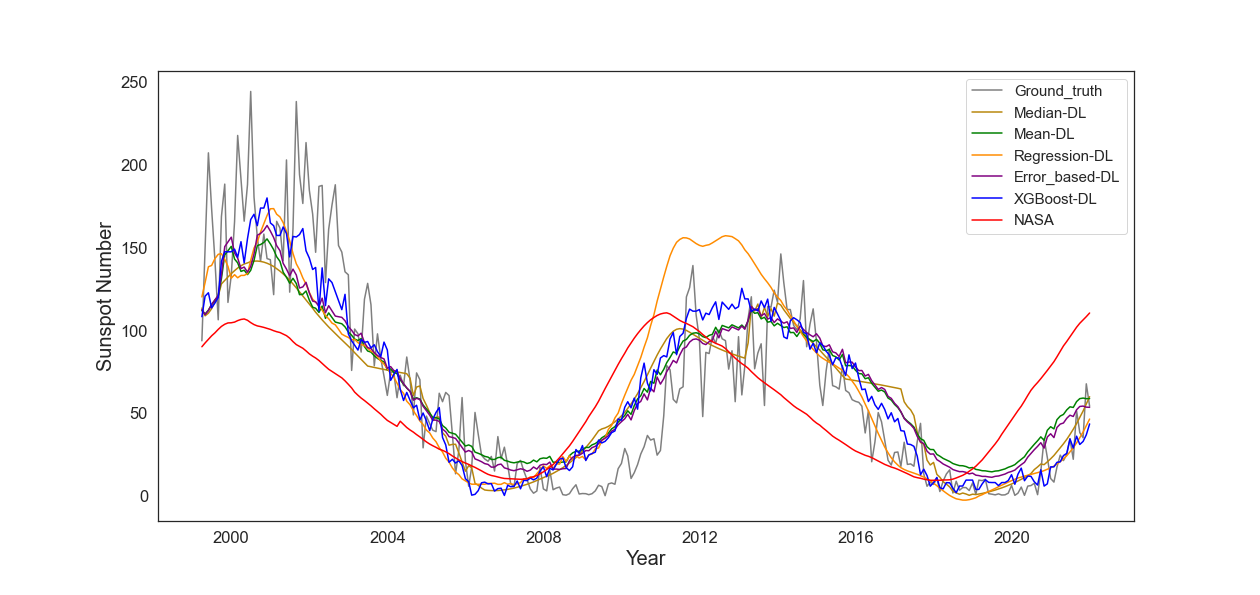}}
\subfigure[Forecasts from ensembles of non-deep learning models.]{
\includegraphics[width=\textwidth]{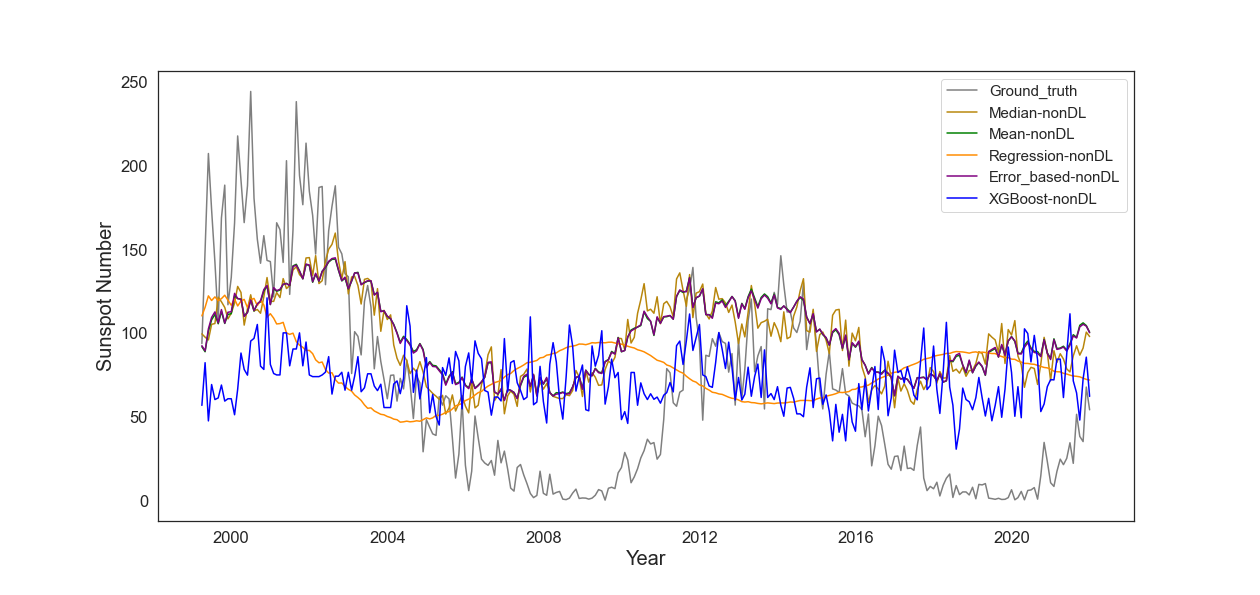}}
\subfigure[Forecasts from ensembles of all the seven base models.]{
\includegraphics[width=\textwidth]{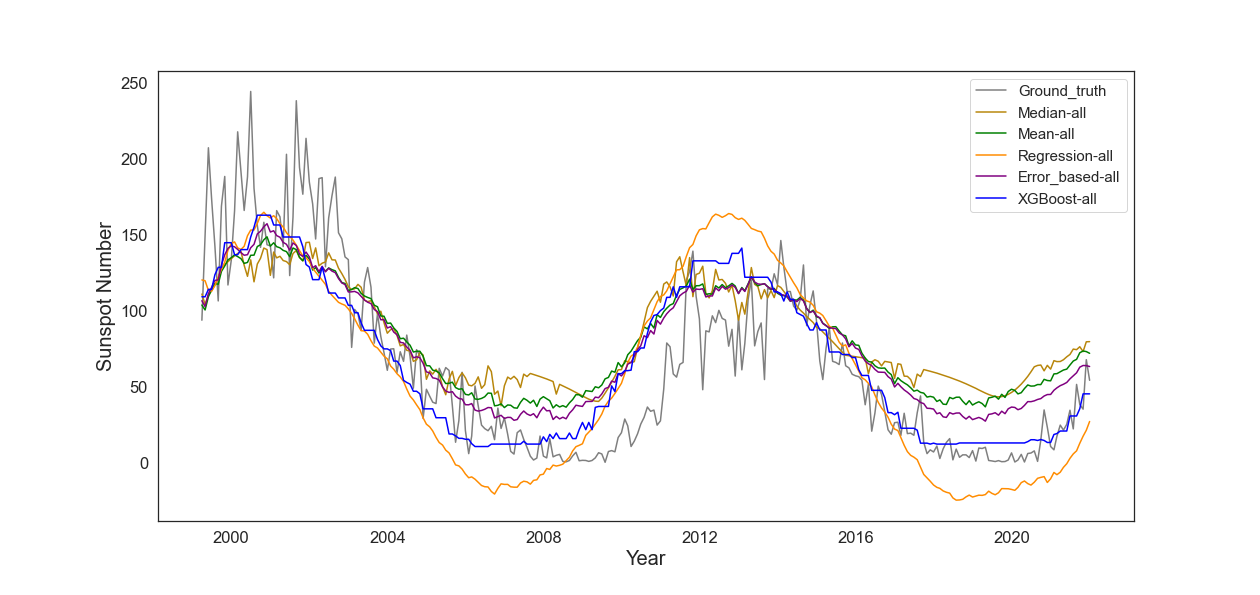}
}
\caption{Forecasts of sunspot numbers from ensemble models for the testing set.}
\label{fig:dl_ensemble_pred}
\end{figure}

In particular, 
the best ensemble model, 
our XGBoost-DL model, i.e.,
the XGBoost ensemble of deep learning models, significantly boosts the performance, with the smallest RMSE of 25.70 and MAE of 19.82 which are $14.05\%$ and $11.32\%$ lower than those of the best base model Informer.  Figure \ref{fig:dl_ensemble_pred} illustrates the forecasts from the three types of ensemble models. All deep-learning ensemble models show strong forecasting abilities in capturing the true trend and seasonality of the series.

{\color{black}\subsubsection{Future forecast}}
We further consider the prediction of future sunspot numbers in current Solar Cycle~25 and the coming Solar Cycle~26. 
{\color{black}
For conciseness of presentation, we only consider XGBoost-DL and NASA for the future forecast, because XGBoost-DL is the best model for the in-sample forecast in Section~\ref{sec: in-sample forecast}, and NASA is the authority in this field, whose forecast serves as the benchmark here due to no ground truth available.}
We fine-tune our XGBoost-DL model using the entire
sunspot number data with a new training and validation split of ratio 7/3 {\color{black}(data from January 1749 to February~1940 for training and the rest for validation)}.

\begin{figure}
\centering
\includegraphics[width=\textwidth]{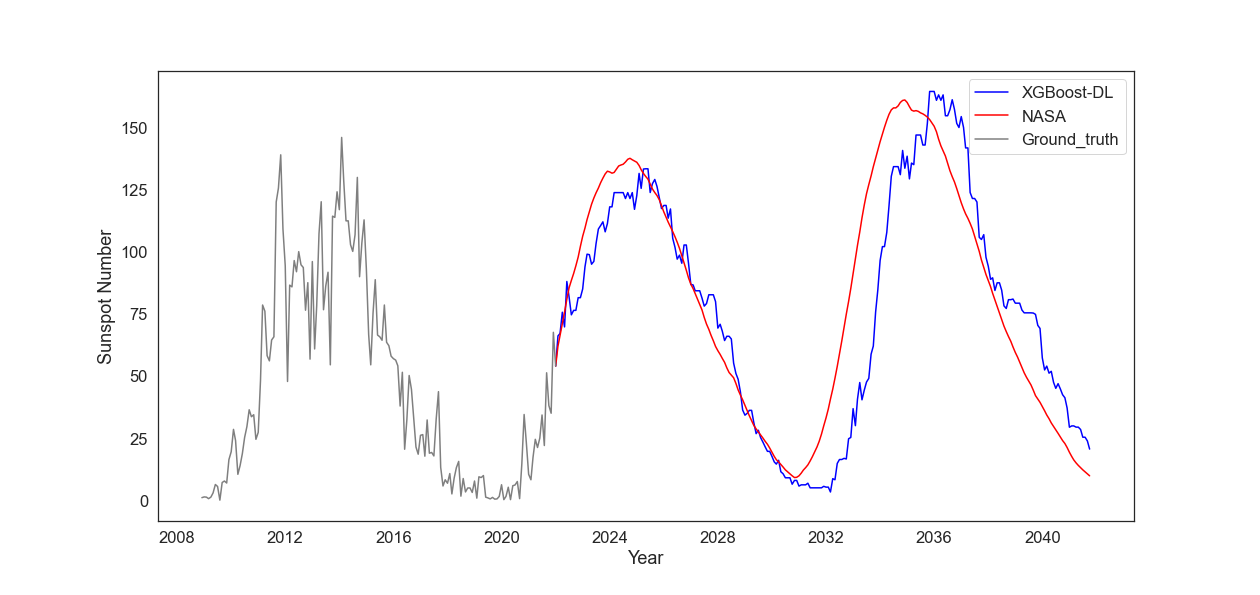}
\caption{Sunspot number forecast from February 2022 to October 2041.}
\label{fig:future_preds}
\end{figure}

Figure \ref{fig:future_preds} shows the forecast from 
our XGBoost-DL model in comparison with NASA's.
 Our forecast indicates that the Solar Cycle 25 will reach a peak sunspot number of 133.47 in May 2025 and Solar Cycle 26 will have a peak number of 164.62 in November 2035. According to our prediction, the two Solar Cycles will be overall stronger than the past Solar Cycle 24. NASA's forecast shows similar but earlier peak values of 137.7 for Solar Cycle 25 in October 2024 and 161.2 for Solar Cycle 26 in December 2034. 
{\color{black}
The similarity of the peaks predicted by XGBoost-DL and NASA moderately shows the plausibility of our XGBoost-DL. But the slightly different time of their predicted peaks is noteworthy to researchers
who use NASA’s prediction,
because it is inferior to XGBoost-DL in the in-sample forecast shown in Section~\ref{sec: in-sample forecast}.}

Although little work has been done on  the sunspot number forecast for Solar Cycle~26, there are a number of recent forecasts for Solar Cycle~25 in the literature with the peak sunspot number ranging from $57.24$ to $228.9$ and occurring between 2022 and 2026.
Our forecast is around the middle of the receptive ranges
of magnitude and time.
\citet{labonville2019dynamo} used a dynamo-based model and forecasted a weak Solar Cycle 25 with a maximum sunspot number of $89^{+29}_{-14}$ in $2025.3^{+0.89}_{-1.05}$. \citet{covas2019neural} applied a feed-forward neural network and obtained a weaker Solar Cycle 25 with the peak sunspot number of $57.24 \pm 16.76$ in about 2022–2023. \citet{han2019decline} predicted a high peak value of $228.9 \pm 40.5$ around $2023.918 \pm 1.64$ years using the Vondrak smoothing method.  \citet{pala2019forecasting} used the LSTM model and predicted a maximum sunspot number of 167.3 in July 2022.
\citet{benson2020forecasting} estimated the peak sunspot number in Solar Cycle 25 to be $106 \pm 19.75$ around March 2025 $\pm$ 1 year by using a combination of the LSTM and WaveNet methods.
\citet{xiong2021forecasting} predicted with multiple regression a peak of 140.2 in March 2024. 
\citet{prasad2022prediction} used a stacked LSTM and predicted the cycle peak with value $171.9 \pm 3.4$ in around August 2023 $\pm$ 2 months.

\bigskip

\section{Conclusion}\label{sec: Conclusion}
We compare three non-deep learning models {\color{black}(SARIMA, Exponential Smoothing, and Prophet)} and four deep learning models {\color{black}(LSTM, GRU, Transformer, and Informer)} to forecast sunspot numbers in this study. The deep learning models outperform the non-deep learning models with lower RMSE and MAE. 
\textcolor{black}{The deep learning models use deep neural networks, which theoretically 
can approximate any
sophisticated nonlinear functions \citep{Haykin2009},
to better model the complex nonlinear temporal dependencies in time series, in contrast to the non-deep learning models that primarily use linear combinations
of time series patterns.
In particular, Transformer and its variant Informer
are superior over LSTM and GRU,
since the former two 
can capture global dependencies in time series
by
adopting
self-attention mechanisms
which ditch the sequential operations
of the recurrent layers in the latter two
and thus provide access to any
part of the sequence.
The best-performing deep learning model, Informer,
enhances Transformer
on long-sequence time series forecasting
by using the ProbSparse self-attention mechanism
and the self-attention distilling operation. 
}

Additionally, five ensemble learning methods 
{\color{black}(via mean, median, the error-based method, linear regression, and XGBoost)} are applied to the forecast results of non-deep learning models, deep learning models, and all base models, separately. Ensemble models
based on deep learning models
have more accurate predictions
than those based on non-deep learning models or all base models.
Our proposed XGBoost-DL model that uses XGBoost to ensemble the four deep learning models  
achieves the best performance among all base and ensemble models as well as NASA's forecast. 
{\color{black}
The outstanding result of
our XGBoost-DL is owing to
its strong two-level nonlinear ensemble 
architecture formed by
decision trees
built upon
deep learning models.}
Our prediction indicates that
Solar Cycles 25 and 26 will be overall stronger than the most recent Solar Cycle 24, and will have the peak sunspot number of 133.47 in May 2025 and 164.62 in November 2035, respectively,
which are similar to but later than 
the peaks forecast by NASA
at 137.7 in October 2024
and 161.2 in December 2034.

{\color{black}There is still room to improve 
our current work.
Our study only considers the point forecasting
without taking into account the prediction intervals of
the forecasted values.
Prediction intervals quantify the forecast uncertainty and are useful in decision making by providing more information to mitigate the risk associated with point forecasts.
Although accurate computation of prediction intervals  
for non-deep learning or deep learning methods
is available, e.g., by using quantile regression \citep{wen2017multi}, the precise estimation method of prediction intervals for ensemble learning is lacking. 
Besides, some researchers \citep{labonville2019dynamo,prasad2022prediction} investigate the sunspot number prediction by only considering the 13-month smoothed monthly sunspot number series instead of its original monthly mean sunspot number series 
which is used in our study and many others \citep{pala2019forecasting,benson2020forecasting,tabassum2020approach}. 
The original monthly data has relatively large fluctuations and thus is more challenging to predict than the smoothed data. 
However, it may be interesting to see whether ensembling the deep learning models trained from smoothed data together with those trained from original monthly data will enhance the forecasting performance. We leave them for our future work.
}

\section*{Disclosure Statement}
No potential conflict of interest was reported by the author(s).

\section*{Funding}
This work was supported by Dr. Hai Shu's startup fund from New York University.

\bibliographystyle{tfcad}
\bibliography{reference}

\begin{thebibliography}{82}
\newcommand{\enquote}[1]{``#1''}
\providecommand{\natexlab}[1]{#1}
\providecommand{\url}[1]{\normalfont{#1}}
\providecommand{\urlprefix}{}

\bibitem[Adhikari and Agrawal(2014)]{adhikari2014performance}
Adhikari, Ratnadip, and RK~Agrawal. 2014. ``Performance evaluation of weights
  selection schemes for linear combination of multiple forecasts.''
  \emph{Artificial Intelligence Review} 42 (4): 529--548.

\bibitem[Allende and Valle(2017)]{allende2017ensemble}
Allende, H{\'e}ctor, and Carlos Valle. 2017. ``Ensemble methods for time series
  forecasting.'' In \emph{Claudio moraga: A passion for multi-valued logic and
  soft computing}, 217--232. Springer.

\bibitem[Arfianti et~al.(2021)]{arfianti2021sunspot}
Arfianti, Unix~Izyah, Dian Candra~Rini Novitasari, Nanang Widodo, Moh
  Hafiyusholeh, and Wika~Dianita Utami. 2021. ``Sunspot Number Prediction Using
  Gated Recurrent Unit (GRU) Algorithm.'' \emph{IJCCS (Indonesian Journal of
  Computing and Cybernetics Systems)} 15 (2): 141--152.

\bibitem[Azc{\'a}rate, Mendoza, and Levi(2016)]{azcarate2016influence}
Azc{\'a}rate, T, B~Mendoza, and JR~Levi. 2016. ``Influence of geomagnetic
  activity and atmospheric pressure on human arterial pressure during the solar
  cycle 24.'' \emph{Advances in Space Research} 58 (10): 2116--2125.

\bibitem[Beltagy, Peters, and Cohan(2020)]{beltagy2020longformer}
Beltagy, Iz, Matthew~E Peters, and Arman Cohan. 2020. ``Longformer: The
  long-document transformer.'' \emph{arXiv preprint arXiv:2004.05150} .

\bibitem[Benson et~al.(2020)]{benson2020forecasting}
Benson, B, WD~Pan, A~Prasad, GA~Gary, and Q~Hu. 2020. ``Forecasting solar cycle
  25 using deep neural networks.'' \emph{Solar Physics} 295: 1--15.

\bibitem[Bhatnagar et~al.(2021)]{bhatnagar2021merlion}
Bhatnagar, Aadyot, Paul Kassianik, Chenghao Liu, Tian Lan, Wenzhuo Yang, Rowan
  Cassius, Doyen Sahoo, et~al. 2021. ``Merlion: A machine learning library for
  time series.'' \emph{arXiv preprint arXiv:2109.09265} .

\bibitem[Box and Jenkins(1976)]{Box1976}
Box, George E.~P., and Gwilym~M. Jenkins. 1976. \emph{Time series analysis:
  forecasting and control}. Revised ed., Holden-Day Series in Time Series
  Analysis. Holden-Day, San Francisco, Calif.-D\"{u}sseldorf-Johannesburg.

\bibitem[Breiman(2001)]{breiman2001random}
Breiman, Leo. 2001. ``Random forests.'' \emph{Machine learning} 45 (1): 5--32.

\bibitem[Breiman et~al.(1984)]{MR726392}
Breiman, Leo, Jerome~H. Friedman, Richard~A. Olshen, and Charles~J. Stone.
  1984. \emph{Classification and regression trees}. Wadsworth
  Statistics/Probability Series. Wadsworth Advanced Books and Software,
  Belmont, CA.

\bibitem[Brockwell and Davis(1991)]{MR1093459}
Brockwell, Peter~J., and Richard~A. Davis. 1991. \emph{Time series: theory and
  methods}. 2nd ed., Springer Series in Statistics. Springer-Verlag, New York.
  \urlprefix\url{https://doi.org/10.1007/978-1-4419-0320-4}.

\bibitem[Brown(1956)]{brown1956exponential}
Brown, R.G. 1956. \emph{Exponential Smoothing for Predicting Demand}. Arthur D.
  Little Inc.
  \urlprefix\url{https://www.industrydocuments.ucsf.edu/tobacco/docs/#id=jzlc0130}.

\bibitem[Cameron and Sch{\"u}ssler(2017)]{cameron2017understanding}
Cameron, RH, and Manfred Sch{\"u}ssler. 2017. ``Understanding solar cycle
  variability.'' \emph{The Astrophysical Journal} 843 (2): 111.

\bibitem[Chang, Chang, and Wu(2018)]{chang2018application}
Chang, Yung-Chia, Kuei-Hu Chang, and Guan-Jhih Wu. 2018. ``Application of
  eXtreme gradient boosting trees in the construction of credit risk assessment
  models for financial institutions.'' \emph{Applied Soft Computing} 73:
  914--920.

\bibitem[Chattopadhyay, Jhajharia, and
  Chattopadhyay(2011)]{chattopadhyay2011trend}
Chattopadhyay, Surajit, Deepak Jhajharia, and Goutami Chattopadhyay. 2011.
  ``Trend estimation and univariate forecast of the sunspot numbers:
  development and comparison of ARMA, ARIMA and autoregressive neural network
  models.'' \emph{Comptes Rendus Geoscience} 343 (7): 433--442.

\bibitem[Chen and Guestrin(2016)]{chen2016xgboost}
Chen, Tianqi, and Carlos Guestrin. 2016. ``Xgboost: A scalable tree boosting
  system.'' In \emph{Proceedings of the 22nd acm sigkdd international
  conference on knowledge discovery and data mining}, 785--794.

\bibitem[Cherkassky and Ma(2004)]{cherkassky2004practical}
Cherkassky, Vladimir, and Yunqian Ma. 2004. ``Practical selection of SVM
  parameters and noise estimation for SVM regression.'' \emph{Neural networks}
  17 (1): 113--126.

\bibitem[Child et~al.(2019)]{child2019generating}
Child, Rewon, Scott Gray, Alec Radford, and Ilya Sutskever. 2019. ``Generating
  long sequences with sparse transformers.'' \emph{arXiv preprint
  arXiv:1904.10509} .

\bibitem[Cho et~al.(2014)]{cho2014}
Cho, Kyunghyun, Bart van Merri{\"e}nboer, Dzmitry Bahdanau, and Yoshua Bengio.
  2014. ``On the Properties of Neural Machine Translation: Encoder{--}Decoder
  Approaches.'' In \emph{Proceedings of {SSST}-8, Eighth Workshop on Syntax,
  Semantics and Structure in Statistical Translation}, Doha, Qatar, 103--111.
  Association for Computational Linguistics.

\bibitem[Chung et~al.(2014)]{chung2014empirical}
Chung, Junyoung, Caglar Gulcehre, Kyunghyun Cho, and Yoshua Bengio. 2014.
  ``Empirical evaluation of gated recurrent neural networks on sequence
  modeling.'' In \emph{NIPS 2014 Workshop on Deep Learning}, ArXiv preprint
  arXiv:1412.3555.

\bibitem[Covas, Peixinho, and Fernandes(2019)]{covas2019neural}
Covas, Eurico, Nuno Peixinho, and Jo{\~a}o Fernandes. 2019. ``Neural network
  forecast of the sunspot butterfly diagram.'' \emph{Solar Physics} 294 (3):
  1--15.

\bibitem[Davis~Jr and Lowell(2004)]{davis2004chaotic}
Davis~Jr, George~E, and Walter~E Lowell. 2004. ``Chaotic solar cycles modulate
  the incidence and severity of mental illness.'' \emph{Medical hypotheses} 62
  (2): 207--214.

\bibitem[Devlin et~al.(2018)]{devlin2018bert}
Devlin, Jacob, Ming-Wei Chang, Kenton Lee, and Kristina Toutanova. 2018.
  ``Bert: Pre-training of deep bidirectional transformers for language
  understanding.'' \emph{arXiv preprint arXiv:1810.04805} .

\bibitem[Dong et~al.(2020)]{dong2020survey}
Dong, Xibin, Zhiwen Yu, Wenming Cao, Yifan Shi, and Qianli Ma. 2020. ``A survey
  on ensemble learning.'' \emph{Frontiers of Computer Science} 14 (2):
  241--258.

\bibitem[Friedman(2001)]{friedman2001greedy}
Friedman, Jerome~H. 2001. ``Greedy function approximation: a gradient boosting
  machine.'' \emph{Annals of statistics} 1189--1232.

\bibitem[Han and Yin(2019)]{han2019decline}
Han, YB, and ZQ~Yin. 2019. ``A decline phase modeling for the prediction of
  solar cycle 25.'' \emph{Solar Physics} 294 (8): 1--14.

\bibitem[Han et~al.(2019)]{han2019review}
Han, Zhongyang, Jun Zhao, Henry Leung, King~Fai Ma, and Wei Wang. 2019. ``A
  review of deep learning models for time series prediction.'' \emph{IEEE
  Sensors Journal} 21 (6): 7833--7848.

\bibitem[Hathaway(2015)]{hathaway2015solar}
Hathaway, David~H. 2015. ``The solar cycle.'' \emph{Living reviews in solar
  physics} 12 (1): 4.

\bibitem[Haykin(2009)]{Haykin2009}
Haykin, Simon. 2009. \emph{Neural networks and learning machines}. 3rd ed.
  Prentice Hall.

\bibitem[Herzen et~al.(2021)]{herzen2021darts}
Herzen, Julien, Francesco L{\"a}ssig, Samuele~Giuliano Piazzetta, Thomas Neuer,
  L{\'e}o Tafti, Guillaume Raille, Tomas Van~Pottelbergh, et~al. 2021. ``Darts:
  User-friendly modern machine learning for time series.'' \emph{arXiv preprint
  arXiv:2110.03224} .

\bibitem[Hiremath(2008)]{hiremath2008prediction}
Hiremath, KM. 2008. ``Prediction of solar cycle 24 and beyond.''
  \emph{Astrophysics and Space Science} 314 (1): 45--49.

\bibitem[Hochreiter(1998)]{hochreiter1998vanishing}
Hochreiter, Sepp. 1998. ``The vanishing gradient problem during learning
  recurrent neural nets and problem solutions.'' \emph{International Journal of
  Uncertainty, Fuzziness and Knowledge-Based Systems} 6 (02): 107--116.

\bibitem[Hochreiter and Schmidhuber(1997)]{hochreiter1997long}
Hochreiter, Sepp, and J{\"u}rgen Schmidhuber. 1997. ``Long short-term memory.''
  \emph{Neural computation} 9 (8): 1735--1780.

\bibitem[Holt(1957)]{Holt1957}
Holt, Charles~C. 1957. \emph{Forecasting seasonals and trends by exponentially
  weighted moving averages}. Office of Naval Research Memorandum 52.

\bibitem[Hyndman and Athanasopoulos(2018)]{hyndman2018forecasting}
Hyndman, R.J., and G.~Athanasopoulos. 2018. \emph{Forecasting: principles and
  practice}. 2nd ed. Melbourne, Australia: OTexts.

\bibitem[Juckett and Rosenberg(1993)]{juckett1993correlation}
Juckett, David~A, and Barnett Rosenberg. 1993. ``Correlation of human longevity
  oscillations with sunspot cycles.'' \emph{Radiation research} 133 (3):
  312--320.

\bibitem[Kaushik et~al.(2020)]{kaushik2020ai}
Kaushik, Shruti, Abhinav Choudhury, Pankaj~Kumar Sheron, Nataraj Dasgupta,
  Sayee Natarajan, Larry~A Pickett, and Varun Dutt. 2020. ``AI in healthcare:
  time-series forecasting using statistical, neural, and ensemble
  architectures.'' \emph{Frontiers in big data} 3: 4.

\bibitem[Kuncheva(2014)]{kuncheva2014combining}
Kuncheva, Ludmila~I. 2014. \emph{Combining pattern classifiers: methods and
  algorithms}. John Wiley \& Sons.

\bibitem[Labonville, Charbonneau, and Lemerle(2019)]{labonville2019dynamo}
Labonville, Francois, Paul Charbonneau, and Alexandre Lemerle. 2019. ``A
  dynamo-based forecast of solar cycle 25.'' \emph{Solar Physics} 294 (6):
  1--14.

\bibitem[Lara-Ben{\'\i}tez, Carranza-Garc{\'\i}a, and
  Riquelme(2021)]{lara2021experimental}
Lara-Ben{\'\i}tez, Pedro, Manuel Carranza-Garc{\'\i}a, and Jos{\'e}~C Riquelme.
  2021. ``An Experimental Review on Deep Learning Architectures for Time Series
  Forecasting.'' \emph{arXiv preprint arXiv:2103.12057} .

\bibitem[Lewandowski(2015)]{lewandowski2015massive}
Lewandowski, K. 2015. ``Massive electricity and communications blackouts on
  Earth as effect of change the Sun activity.'' \emph{Journal of Polish Safety
  and Reliability Association} 6 (3): 91--98.

\bibitem[Li and Racine(2007)]{li2007nonparametric}
Li, Qi, and Jeffrey~Scott Racine. 2007. \emph{Nonparametric econometrics:
  theory and practice}. Princeton University Press.

\bibitem[Li et~al.(2019)]{li2019enhancing}
Li, Shiyang, Xiaoyong Jin, Yao Xuan, Xiyou Zhou, Wenhu Chen, Yu-Xiang Wang, and
  Xifeng Yan. 2019. ``Enhancing the locality and breaking the memory bottleneck
  of transformer on time series forecasting.'' \emph{Advances in Neural
  Information Processing Systems} 32: 5243--5253.

\bibitem[Liaw et~al.(2018)]{liaw2018tune}
Liaw, Richard, Eric Liang, Robert Nishihara, Philipp Moritz, Joseph~E Gonzalez,
  and Ion Stoica. 2018. ``Tune: A Research Platform for Distributed Model
  Selection and Training.'' \emph{arXiv preprint arXiv:1807.05118} .

\bibitem[Lidiema(2017)]{lidiema2017modelling}
Lidiema, Caspah. 2017. ``Modelling and Forecasting Inflation Rate in Kenya
  Using SARIMA and Holt-Winters Triple Exponential Smoothing.'' \emph{American
  Journal of Theoretical and Applied Statistics} 6 (3): 161--169.

\bibitem[Liu et~al.(2020)]{liu2020forecast}
Liu, Huan, Chenxi Li, Yingqi Shao, Xin Zhang, Zhao Zhai, Xing Wang, Xinye Qi,
  et~al. 2020. ``Forecast of the trend in incidence of acute hemorrhagic
  conjunctivitis in China from 2011--2019 using the Seasonal Autoregressive
  Integrated Moving Average (SARIMA) and Exponential Smoothing (ETS) models.''
  \emph{Journal of infection and public health} 13 (2): 287--294.

\bibitem[Liu et~al.(2019)]{liu2019roberta}
Liu, Yinhan, Myle Ott, Naman Goyal, Jingfei Du, Mandar Joshi, Danqi Chen, Omer
  Levy, Mike Lewis, Luke Zettlemoyer, and Veselin Stoyanov. 2019. ``Roberta: A
  robustly optimized bert pretraining approach.'' \emph{arXiv preprint
  arXiv:1907.11692} .

\bibitem[Lybekk et~al.(2012)]{lybekk2012solar}
Lybekk, Bj{\o}rn, Arne Pedersen, Stein Haaland, Knut Svenes, Andrew~N
  Fazakerley, A~Masson, MGGT Taylor, and J-G Trotignon. 2012. ``Solar cycle
  variations of the Cluster spacecraft potential and its use for electron
  density estimations.'' \emph{Journal of Geophysical Research: Space Physics}
  117 (A1).

\bibitem[Moritz et~al.(2018)]{moritz2018ray}
Moritz, Philipp, Robert Nishihara, Stephanie Wang, Alexey Tumanov, Richard
  Liaw, Eric Liang, Melih Elibol, et~al. 2018. ``Ray: A distributed framework
  for emerging $\{$AI$\}$ applications.'' In \emph{13th $\{$USENIX$\}$
  Symposium on Operating Systems Design and Implementation ($\{$OSDI$\}$ 18)},
  561--577.

\bibitem[Ogunleye and Wang(2019)]{ogunleye2019xgboost}
Ogunleye, Adeola, and Qing-Guo Wang. 2019. ``XGBoost model for chronic kidney
  disease diagnosis.'' \emph{IEEE/ACM transactions on computational biology and
  bioinformatics} 17 (6): 2131--2140.

\bibitem[Oliveira and Torgo(2015)]{oliveira2015ensembles}
Oliveira, Mariana, and Luis Torgo. 2015. ``Ensembles for time series
  forecasting.'' In \emph{Asian Conference on Machine Learning}, 360--370.
  PMLR.

\bibitem[Pala and Atici(2019)]{pala2019forecasting}
Pala, Zeydin, and Ramazan Atici. 2019. ``Forecasting sunspot time series using
  deep learning methods.'' \emph{Solar Physics} 294 (5): 1--14.

\bibitem[Pan(2018)]{pan2018application}
Pan, Bingyue. 2018. ``Application of XGBoost algorithm in hourly PM2. 5
  concentration prediction.'' In \emph{IOP conference series: earth and
  environmental science}, Vol. 113, 012127. IOP publishing.

\bibitem[Paszke et~al.(2019)]{paszke2019pytorch}
Paszke, Adam, Sam Gross, Francisco Massa, Adam Lerer, James Bradbury, Gregory
  Chanan, Trevor Killeen, et~al. 2019. ``Pytorch: An imperative style,
  high-performance deep learning library.'' \emph{Advances in neural
  information processing systems} 32: 8026--8037.

\bibitem[Penrose(1956)]{penrose1956best}
Penrose, Roger. 1956. ``On best approximate solutions of linear matrix
  equations.'' In \emph{Mathematical Proceedings of the Cambridge Philosophical
  Society}, Vol.~52, 17--19. Cambridge University Press.

\bibitem[Prasad et~al.(2022)]{prasad2022prediction}
Prasad, Amrita, Soumya Roy, Arindam Sarkar, Subhash~Chandra Panja, and
  Sankar~Narayan Patra. 2022. ``Prediction of solar cycle 25 using deep
  learning based long short-term memory forecasting technique.'' \emph{Advances
  in Space Research} 69 (1): 798--813.

\bibitem[Pulkkinen(2007)]{pulkkinen2007space}
Pulkkinen, Tuija. 2007. ``Space weather: terrestrial perspective.''
  \emph{Living Reviews in Solar Physics} 4 (1): 1--60.

\bibitem[Qiu et~al.(2014)]{qiu2014ensemble}
Qiu, Xueheng, Le~Zhang, Ye~Ren, Ponnuthurai~N Suganthan, and Gehan Amaratunga.
  2014. ``Ensemble deep learning for regression and time series forecasting.''
  In \emph{2014 IEEE symposium on computational intelligence in ensemble
  learning (CIEL)}, 1--6. IEEE.

\bibitem[Qu(2016)]{qu2016sunspot}
Qu, Jiangwen. 2016. ``Is sunspot activity a factor in influenza pandemics?''
  \emph{Reviews in medical virology} 26 (5): 309--313.

\bibitem[Rabbani et~al.(2021)]{rabbani2021comparison}
Rabbani, Muhammad Babar~Ali, Muhammad~Ali Musarat, Wesam~Salah Alaloul,
  Muhammad~Shoaib Rabbani, Ahsen Maqsoom, Saba Ayub, Hamna Bukhari, and
  Muhammad Altaf. 2021. ``A Comparison Between Seasonal Autoregressive
  Integrated Moving Average (SARIMA) and Exponential Smoothing (ES) Based on
  Time Series Model for Forecasting Road Accidents.'' \emph{Arabian Journal for
  Science and Engineering} 46 (11): 11113--11138.

\bibitem[Ruohoniemi and Greenwald(1997)]{ruohoniemi1997rates}
Ruohoniemi, JM, and RA~Greenwald. 1997. ``Rates of scattering occurrence in
  routine HF radar observations during solar cycle maximum.'' \emph{Radio
  Science} 32 (3): 1051--1070.

\bibitem[Sagi and Rokach(2018)]{sagi2018ensemble}
Sagi, Omer, and Lior Rokach. 2018. ``Ensemble learning: A survey.'' \emph{Wiley
  Interdisciplinary Reviews: Data Mining and Knowledge Discovery} 8 (4): e1249.

\bibitem[Suggs(2017)]{suggs2017msfc}
Suggs, Ronnie~J. 2017. ``The MSFC Solar Activity Future Estimation (MSAFE)
  Model.'' In \emph{The Applied Space Environments Conference}, M17--6038.

\bibitem[Tabassum, Rabbani, and Omar(2020)]{tabassum2020approach}
Tabassum, Anika, Masud Rabbani, and Saad~Bin Omar. 2020. ``An Approach to Study
  on MA, ES, AR for Sunspot Number (SN) Prediction and to Forecast SN with
  Seasonal Variations Along with Trend Component of Time Series Analysis Using
  Moving Average (MA) and Exponential Smoothing (ES).'' In \emph{Advances in
  Electrical and Computer Technologies}, 373--386. Springer.

\bibitem[Taylor and Letham(2018)]{taylor2018forecasting}
Taylor, Sean~J, and Benjamin Letham. 2018. ``Forecasting at scale.'' \emph{The
  American Statistician} 72 (1): 37--45.

\bibitem[Torres et~al.(2021)]{torres2021deep}
Torres, Jos{\'e}~F, Dalil Hadjout, Abderrazak Sebaa, Francisco
  Mart{\'\i}nez-{\'A}lvarez, and Alicia Troncoso. 2021. ``Deep Learning for
  Time Series Forecasting: A Survey.'' \emph{Big Data} 9 (1): 3--21.

\bibitem[Tsui et~al.(1995)]{tsui1995recurrent}
Tsui, Fu-Chiang, Mingui Sun, Ching-Chung Li, and Robert~J Sclabassi. 1995.
  ``Recurrent neural networks and discrete wavelet transform for time series
  modeling and prediction.'' In \emph{1995 International Conference on
  Acoustics, Speech, and Signal Processing}, Vol.~5, 3359--3362. IEEE.

\bibitem[Usoskin(2017)]{usoskin2017history}
Usoskin, Ilya~G. 2017. ``A history of solar activity over millennia.''
  \emph{Living Reviews in Solar Physics} 14 (1): 1--97.

\bibitem[Van~Rossum and Drake(2009)]{10.5555/1593511}
Van~Rossum, Guido, and Fred~L. Drake. 2009. \emph{Python 3 Reference Manual}.
  Scotts Valley, CA: CreateSpace.

\bibitem[Vaswani et~al.(2017)]{vaswani2017attention}
Vaswani, Ashish, Noam Shazeer, Niki Parmar, Jakob Uszkoreit, Llion Jones,
  Aidan~N Gomez, {\L}ukasz Kaiser, and Illia Polosukhin. 2017. ``Attention is
  all you need.'' In \emph{Advances in neural information processing systems},
  5998--6008.

\bibitem[Walterscheid(1989)]{walterscheid1989solar}
Walterscheid, RL. 1989. ``Solar cycle effects on the upper
  atmosphere-Implications for satellite drag.'' \emph{Journal of spacecraft and
  rockets} 26 (6): 439--444.

\bibitem[Wen et~al.(2020)]{wen2020time}
Wen, Qingsong, Liang Sun, Fan Yang, Xiaomin Song, Jingkun Gao, Xue Wang, and
  Huan Xu. 2020. ``Time series data augmentation for deep learning: A survey.''
  \emph{arXiv preprint arXiv:2002.12478} .

\bibitem[Wen et~al.(2017)]{wen2017multi}
Wen, Ruofeng, Kari Torkkola, Balakrishnan Narayanaswamy, and Dhruv Madeka.
  2017. ``A multi-horizon quantile recurrent forecaster.'' \emph{arXiv preprint
  arXiv:1711.11053} .

\bibitem[Wichard and Ogorzalek(2004)]{wichard2004time}
Wichard, Jorg~D, and Maciej Ogorzalek. 2004. ``Time series prediction with
  ensemble models.'' In \emph{2004 IEEE International Joint Conference on
  Neural Networks (IEEE Cat. No. 04CH37541)}, Vol.~2, 1625--1630. IEEE.

\bibitem[Winters(1960)]{winters1960forecasting}
Winters, Peter~R. 1960. ``Forecasting sales by exponentially weighted moving
  averages.'' \emph{Management science} 6 (3): 324--342.

\bibitem[Wu et~al.(2020)]{wu2020deep}
Wu, Neo, Bradley Green, Xue Ben, and Shawn O'Banion. 2020. ``Deep transformer
  models for time series forecasting: The influenza prevalence case.''
  \emph{arXiv preprint arXiv:2001.08317} .

\bibitem[Xiong et~al.(2021)]{xiong2021forecasting}
Xiong, Yating, Jianyong Lu, Kai Zhao, Meng Sun, and Yang Gao. 2021.
  ``Forecasting Solar Cycle 25 using comprehensive precursor combination and
  multiple regression technique.'' \emph{Monthly Notices of the Royal
  Astronomical Society} 505 (1): 1046--1052.

\bibitem[Xu et~al.(2008)]{xu2008long}
Xu, Tong, Jian Wu, Zhen-Sen Wu, and Qiang Li. 2008. ``Long-term sunspot number
  prediction based on EMD analysis and AR model.'' \emph{Chinese journal of
  astronomy and astrophysics} 8 (3): 337.

\bibitem[Yang et~al.(2019)]{yang2019xlnet}
Yang, Zhilin, Zihang Dai, Yiming Yang, Jaime Carbonell, Russ~R Salakhutdinov,
  and Quoc~V Le. 2019. ``Xlnet: Generalized autoregressive pretraining for
  language understanding.'' \emph{Advances in neural information processing
  systems} 3232.

\bibitem[Zhang and Man(1998)]{zhang1998time}
Zhang, Jun, and Kim-Fung Man. 1998. ``Time series prediction using RNN in
  multi-dimension embedding phase space.'' In \emph{SMC'98 Conference
  Proceedings. 1998 IEEE International Conference on Systems, Man, and
  Cybernetics (Cat. No. 98CH36218)}, Vol.~2, 1868--1873. IEEE.

\bibitem[Zhong et~al.(2018)]{zhong2018xgbfemf}
Zhong, Jiancheng, Yusui Sun, Wei Peng, Minzhu Xie, Jiahong Yang, and Xiwei
  Tang. 2018. ``XGBFEMF: an XGBoost-based framework for essential protein
  prediction.'' \emph{IEEE transactions on nanobioscience} 17 (3): 243--250.

\bibitem[Zhou et~al.(2021)]{zhou2021informer}
Zhou, Haoyi, Shanghang Zhang, Jieqi Peng, Shuai Zhang, Jianxin Li, Hui Xiong,
  and Wancai Zhang. 2021. ``Informer: Beyond Efficient Transformer for Long
  Sequence Time-Series Forecasting.'' In \emph{The Thirty-Fifth {AAAI}
  Conference on Artificial Intelligence, {AAAI} 2021, Virtual Conference},
  Vol.~35, 11106--11115. {AAAI} Press.

\end{thebibliography}

\end{document}